\documentclass[12pt]{iopart}
\usepackage{graphicx} 
\usepackage{dcolumn} 
\usepackage{amssymb}
\usepackage{iopams}

\begin{document}

\title[Schwarzschild Black Hole Threaded by a Cosmic String]{Renormalized Vacuum Polarization on the Horizon of a Schwarzschild Black Hole Threaded by a Cosmic String}
\author{Adrian C. Ottewill and Peter Taylor}
\address{School of Mathematical Sciences \& Complex and Adaptive Systems Laboratory, University College Dublin, Belfield, Dublin 4, Ireland}
\eads{\mailto{adrian.ottewill@ucd.ie} and \mailto{peter.taylor@ucd.ie}}
\date{\today}
\begin{abstract}
We obtain an analytic expression for the vacuum polarization on the horizon of the Schwarzschild black hole threaded by an infinite cosmic string. This calculation relies on a generalized Heine  
identity for Legendre functions which we derive without using specific properties of the Legendre functions themselves.
\end{abstract}
\maketitle


\section{Introduction}
There is a long and successful history of  calculating the renormalized vacuum polarization and stress-energy tensor on spherically-symmetric black holes (see \cite{Candelas:1980zt,Candelas:1984pg,Anderson:1989vg,JensenOttewill, JMO1, JMO2, Anderson:1990jh,Winstanley:2007}), however, in the astrophysically significant case of the Kerr-Newman black hole, such calculations  have proved intractable. This is due to the fact that all the calculations in the spherically symmetric case have relied heavily on specific properties of the Legendre functions (mainly the Legendre Addition Theorem and the Heine Identity) and also the applicability of the Watson-Sommerfeld Formula to obtain a mode-sum expression that is numerically tractable. In the axially symmetric case, however, addition theorems for the angular functions do not exist nor do we have an analog of the Heine identity for the angular functions that arise. We will show, in the framework of the Schwarzschild black hole threaded by a cosmic string, that we can proceed in the axially symmetric case by obtaining useful summation formulae based on the Hadamard structure of the Green function without using specific properties of the mode-functions themselves. In this paper we will
use this approach to obtain an analytic expression for the renormalized vacuum polarization on the horizon of such a space-time.

It should be noted that the result presented here is a generalization of a result given by Davies and Sahni \cite{DaviesSahni} where the authors calculate the vacuum polarization for a restricted and discrete set of values for the azimuthal deficit.


\section{The Green's Function}
 The cosmic string is modeled by introducing an azimuthal deficit parameter, $\alpha$, into the standard Schwarzschild metric. We
may describe this in  coordinates $(t, r, \theta, \tilde{\phi})$, where  $\tilde{\phi}$ is periodic with period  $2\pi \alpha$, so we may take $\tilde{\phi}\in[0,  2\pi \alpha)$, in which the line element is given by
\begin{equation}
\fl
{\rmd}s^2 = -(1-2M/r) {\rmd}t^2 +(1-2M/r)^{-1} {\rmd}r^2 +r^2 {\rmd} \theta^2 +  r^{2} \sin^{2}\theta {\rmd} \tilde{\phi}^{2}.
\end{equation}
For GUT scale cosmic strings $\alpha = 1- 4 \mu$ where $\mu$ is the mass per unit length of the string so we shall assume $0 < \alpha \le 1$.
We can alternatively define a new azimuthal coordinate by
\begin{equation}
\tilde{\phi} = \alpha \phi
\end{equation}
so that  $\phi$ is periodic with period  $2\pi$ and we may take $\phi \in [0, 2 \pi)$. In coordinates $(t, r, \theta, \phi)$,  the line element is given by
\begin{equation}
\fl
{\rmd}s^2 = -(1-2M/r) {\rmd}t^2 +(1-2M/r)^{-1} {\rmd}r^2 +r^2 {\rmd} \theta^2 + \alpha^{2} r^{2} \sin^{2}\theta {\rmd} \phi^{2}.
\end{equation}

We shall consider a massless, minimally coupled scalar field, $\varphi$, in the Hartle-Hawking vacuum state. 
Since this is a thermal state, it is convenient to work with the Euclidean Green's function, performing a Wick rotation of the
 temporal coordinate $t\rightarrow -i \tau$ and eliminating the conical singularity at $r=2M$ by making $\tau$ periodic with period
$2\pi/\kappa$ where $\kappa=1/(4M)$ is  the surface gravity of the black hole. 
The massless, minimally coupled scalar field, satisfies the homogenous wave-equation
\begin{equation}
\Box \varphi(\tau,r,\theta,\phi)=0,
\end{equation}
which can be solved by a separation of variables by writing
\begin{equation}
\varphi(\tau,r,\theta,\phi)\sim \rme^{i n \kappa \tau+i m\phi} P(\theta)R(r)
\end{equation}
where $P(\theta)$ is regular and satisfies 
\begin{equation}
\label{eq:legendre}
\Big\{\frac{1}{\sin\theta} \frac{{\rmd}}{{\rmd} \theta}\Big(\sin\theta\frac{{\rmd}}{{\rmd} \theta}\Big) -\frac{m^{2}}{\alpha^{2}\sin^{2}\theta}+\lambda(\lambda+1)\Big\}P(\theta)=0
\end{equation}
while $R(r)$ satisfies
\begin{equation}
\label{eq:radialhomogeneous}
\Big\{\frac{{\rmd}}{{\rmd} r}(r^2-2Mr)\frac{{\rmd}}{{\rmd}r} - \lambda(\lambda+1)-\frac{n^2 \kappa^2 r^4}{r^2-2Mr}\Big\}R(r)= 0.
\end{equation}
 The $\lambda(\lambda+1)$ term arises as the separation constant. The choice of $\lambda$ is arbitrary for $\varphi$ to satisfy the wave equation but requires a specific choice in order for the mode-function $P(\theta)$ to satisfy the boundary conditions of regularity on the poles. In the Schwarzschild case (no cosmic string, $\alpha=1$), regularity on the poles means that $\lambda=l$, i.e the separation constant is $l(l+1)$. In the cosmic string case, the appropriate choice of $\lambda$ that guarantees regularity of the angular functions on the poles is
\begin{equation}
\label{eq:lambda}
\lambda=l-|m|+|m|/\alpha.
\end{equation}  
With this choice of $\lambda$, the angular function is the Legendre function of both non-integer order and non-integer degree,\textit{viz.},
\begin{equation}
P(\theta)=P_{\lambda}^{-|m|/\alpha}(\cos\theta).
\end{equation}
It can be shown that these angular functions satisfy the following normalization condition,
\begin{eqnarray}
\label{eq:norm}
&\int_{-1}^{1}P_{l-|m|+|m|/\alpha}^{-|m|/\alpha}(\cos\theta) P_{l'-|m|+|m|/\alpha}^{-|m|/\alpha}(\cos\theta) d(\cos\theta) 
=\nonumber\\
&\qquad\qquad\qquad\frac{2}{(2\lambda+1)}\frac{\Gamma(\lambda-|m|/\alpha+1)}{\Gamma(\lambda+|m|/\alpha+1)}\delta_{ll'}. 
\end{eqnarray}

The periodicity of the Green's function with respect to $(\tau-\tau')$ and $(\phi-\phi')$ with periodicity $2\pi/\kappa$ and $2\pi$, respectively, combined with Eq.(\ref{eq:norm}) allow us to write the mode-sum expression for the Green's function as
\begin{eqnarray}
\label{eq:greensfn}
\fl
G(x,x')=\frac{\kappa}{8\pi^{2}}\sum_{n=-\infty}^{\infty} {\rme}^{i n \kappa (\tau-\tau')}&\sum_{m=-\infty}^{\infty} {\rme}^{im(\phi-\phi')} \sum_{l=|m|}^{\infty} (2\lambda+1)\frac{\Gamma(\lambda+|m|/\alpha+1)}{\Gamma(\lambda-|m|/\alpha+1)}  \nonumber\\
 &\qquad P_{\lambda}^{-|m|/\alpha}(\cos\theta) P_{\lambda}^{-|m|/\alpha}(\cos\theta')\chi_{n\lambda}(r,r'),
\end{eqnarray}
where $\chi_{n\lambda}(r,r')$ satisfies the inhomogeneous equation,
\begin{equation}
\label{eq:radialr}
\fl
\Big\{\frac{{\rmd}}{{\rmd} r}(r^2-2Mr)\frac{{\rmd}}{{\rmd}r} - \lambda(\lambda+1)-\frac{n^2 \kappa^2 r^4}{r^2-2Mr}\Big\}\chi_{n\lambda}(r,r')=-\frac{1}{\alpha}\delta(r-r').\nonumber\\
\end{equation}

It is convenient to write the radial equation in terms of a new radial variable
$\eta=r/M-1$,
 the radial equation then reads 
\begin{eqnarray}
\label{eq:radial}
\fl
\Big\{\frac{{\rmd}}{{\rmd}\eta}\Big((\eta^{2}-1)\frac{{\rmd}}{{\rmd}\eta}\Big)-\lambda(\lambda+1)-\frac{n^{2}(1+\eta)^{4}}{16(\eta^{2}-1)}\Big\}\chi_{n\lambda}(\eta,\eta')=-\frac{1}{\alpha M}\delta(\eta-\eta'),
\end{eqnarray}
where we have used the fact that $\kappa=1/4M$. For $n=0$, the two solutions of the homogeneous equation are the Legendre functions of the first and second kind. For $n\ne 0$, the homogeneous equation cannot be solved in terms of known functions and must be solved numerically. We denote the two solutions that are regular on the horizon and infinity (or some outer boundary) by $p_{\lambda}^{|n|}(\eta)$ and $q_{\lambda}^{|n|}(\eta)$, respectively. A near-horizon Frobenius analysis for $n \neq 0$ shows that the indicial exponent is $\pm |n|/2$, and so we have the following asymptotic forms:
\begin{equation}
\label{eq:asymp}
\eqalign{
 p_{\lambda}^{|n|}(\eta)\sim (\eta-1)^{|n|/2} \qquad\qquad &\eta\rightarrow 1, \cr
q_{\lambda}^{|n|}(\eta) \sim (\eta-1)^{-|n|/2} &\eta \rightarrow 1.}
\end{equation}
Defining the normalizations by these asymptotic forms and using the Wronskian conditions one can obtain the appropriate normalization of the Green's function:
 \begin{eqnarray}
 \label{eq:chi}
 \chi_{n\lambda}(\eta,\eta') = 
\cases{
\displaystyle{\frac{1}{\alpha M}} P_{\lambda}(\eta_{<})Q_{\lambda}(\eta_{>})&$n=0$, \\
\displaystyle \frac{1}{2|n|\alpha M} p_{\lambda}^{|n|}(\eta_{<}) q_{\lambda}^{|n|}(\eta_{>})\qquad&$n\neq 0$.}
 \end{eqnarray}


\section{Renormalization}
 From the asymptotic forms (\ref{eq:asymp}), it is clear that taking one point on the horizon (with the other outside) means that all the $n\neq 0$ modes vanish. The unrenormalized Green's function then reduces to
\begin{eqnarray}
\label{eq:greensfnhorizon}
\fl
G(x,x')=\frac{1}{32 \pi^{2} M^{2} \alpha}&\sum_{m=-\infty}^{\infty} {\rme}^{im(\phi-\phi')} \sum_{l=|m|}^{\infty} (2 \lambda+1) \frac{\Gamma(\lambda+|m|/\alpha+1)}{\Gamma(\lambda-|m|/\alpha+1)} \nonumber\\
& \qquad P^{-|m|/\alpha}_{\lambda}(\cos\theta) P^{-|m|/\alpha}_{\lambda}(\cos\theta') Q_{\lambda}(\eta).
\end{eqnarray}
In the absence of the cosmic string, i.e. $\alpha = 1$, this sum can be done using a combination of the Legendre Addition Theorem and the Heine Identity, yielding a closed form expression~\cite{Candelas:1980zt}, that,
of course, diverges in the coincidence limit. 

To renormalize we employ the Christensen-DeWitt\cite{Christensen:1976vb} point-separation method subtracting the geometrical singular
part of the Green's function for small separations in the coordinates. This method relies on the universal Hadamard singularity structure  of the Green's function which ensures that once we have subtracted the geometric singular part, `$U/\sigma + V \ln \sigma$', the coincidence limit of the remainder `$W$' is finite. To calculate $\langle \hat{\varphi}^2 \rangle_{ren}$ in the massless Ricci-flat case, the only subtraction term needed is
\begin{equation}
G_{sing}(x,x') = \frac{1}{4\pi^{2} s^2}
\end{equation}
where $s$ is the geodesic distance between $x$ and $x'$. Having already taken one point on the horizon
we choose to separate radially, placing $x'$ at $2M+\epsilon$; the geodesic distance is then given by
\begin{eqnarray}
\fl
 s(2M,2M+\epsilon)&=&\int_{2 M}^{2M+\epsilon} \frac{\rmd r'}{(1-2M/r')^{1/2}}  =  (2M\epsilon)^{1/2} \bigg[2+1/3(\epsilon/2M) +O(\epsilon^2)\bigg].
 \end{eqnarray}
Note that this behaves like $(\Delta x)^{1/2}$ rather that the usual $\Delta x$ since the metric is singular at 
$r=2M$ so $g^{r'r'}=O(\epsilon)$, but this is simply a coordinate singularity and in no way affects the validity of the geometrical subtraction.  The subtraction terms up to $O(\epsilon)$ are
 \begin{equation}
 \label{eq:gdiv}
 G_{sing}  =\frac{1}{32 \pi^{2} M \epsilon} - \frac{1}{192 \pi^{2} M^{2}} + O(\epsilon).
 \end{equation}
 
Taking the partial coincidence limit $\phi \rightarrow \phi'$, $\theta\rightarrow \theta'$ in (\ref{eq:greensfnhorizon}), subtracting the renormalization terms (\ref{eq:gdiv}) and taking the limit as $\eta \rightarrow 1$ ($\epsilon\rightarrow 0$) gives us the renormalized expression for the vacuum polarization. However, we now face our fundamental challenge as the limit cannot be taken in a meaningful way in the forms given. In order to take the limit, we must either invent a way to write the divergent terms as an appropriate mode-sum and do a mode-by-mode subtraction or alternatively, we must attempt to perform the sum in (\ref{eq:greensfnhorizon}) so that we can write the Green's function in closed form. In \cite{CSHorizon} we follow the former route to obtain numerical values of $\langle \hat{\varphi}^2 \rangle_{ren}$ outside our black hole. However, when possible, the latter is clearly preferred since it can give us a simple closed-form solution for the renormalized vacuum polarization. In the next section, we derive a formula that will allow us to perform the sum (\ref{eq:greensfnhorizon}).


\section{Generalized Heine Identity}
We recall that for real arguments, the Heine Identity for the Legendre Functions is \cite{Erdelyi}
\begin{equation}
\label{eq:heine}
\sum_{l=0}^{\infty} (2 l+1) P_{l}(\Psi)Q_{l}(\zeta)=\frac{1}{(\zeta-\Psi)},
\end{equation}
valid for $|\Psi|<|\zeta|$. Writing $\Psi=\cos\gamma=\cos\theta\cos\theta'+\sin\theta\sin\theta'\cos(\phi-\phi')$, and using the Legendre Addition Theorem,
\begin{equation}
P_{l}(\cos\gamma) = \sum_{m=-l}^{l} \rme^{im (\phi-\phi')} \frac{(l-m)!}{(l+m)!} P_{l}^{m}(\cos\theta)P_{l}^{m}(\cos\theta'),
\end{equation}
we can re-write the Heine Identity as
\begin{equation}
\label{eq:heinenew}
\fl
\sum_{l=0}^{\infty}\sum_{m=-l}^{l} \rme^{im (\phi-\phi')}(2 l+1)\frac{(l-m)!}{(l+m)!} P_{l}^{m}(\cos\theta)P_{l}^{m}(\cos\theta') Q_{l}(\zeta)=\frac{1}{(\zeta-\cos\gamma)}.
\end{equation}
Below we shall generalize this form of the Heine Identity to the non-integer Legendre functions that arise in the mode-sum of a cosmic string space-time. 
It is this form rather than a generalization of (\ref{eq:heine}) that is useful to us since the fundamentaly axially-symmetric nature of the Green's function for $\alpha\neq 1$ means that its radial part  will always depend on $m$ (through $\lambda$), and therefore the $m$-sum over the angular functions can never be done independently of the radial part.

In the Appendix, we have derived other identities involving Legendre Functions of non-integer order and/or degree. We have done so since, although they are not directly useful here, they turn out to be useful in other contexts, for example, determining the Hadamard singularity structure of a 3D Green's function on a cosmic string black hole space-time \cite{CSHorizon}.

Returning to the problem at hand, we want to construct a generalization of (\ref{eq:heinenew}). In addition, we would prefer to do this in such a way that the derivation does not rely on specific properties of the Legendre functions themselves, since
ultimately we wish to apply the same techniques in Kerr space-time. There is a very natural way to do this by equating equivalent expressions for the same Green's function on a conveniently chosen space-time. 
In our case, the most appropriate case to consider is that of  an infinite cosmic string in otherwise flat space-time. Of course, normally in this case one would exploit translational invariance along the 
string and work in cylindrical polar coordinates but for our purposes we choose to work in spherical polar coordinates, so
\begin{equation}
{\rmd}s^{2} = {\rmd}\tau^{2} +{\rmd}r^{2} +r^{2} {\rmd}\theta^{2} +\alpha^{2} r^{2} \sin^{2}\theta {\rmd}\phi^{2},
\end{equation}
where $\phi$ is periodic with period  $2\pi$.
The Green's function for a minimally coupled scalar field on this space-time has a closed-form solution,
which can be obtained, for example, by separating in cylindrical polar coordinates~\cite{Smith}
\begin{equation}
\label{eq:gclosed}
G(x,x') = \frac{1}{8\pi^{2}\alpha} \frac{\sinh(\chi/\alpha)}{\rho\rho' \sinh\chi(\cosh(\chi/\alpha)-\cos(\phi-\phi'))},
\end{equation}
where
\begin{equation}
\cosh\chi = \frac{(\tau-\tau')^{2}+r^2 +r'^2-2r r' \cos\theta\cos\theta'}{2\rho\rho'}
\end{equation}
and $\rho=r\sin\theta$ is the cylindrical polar coordinate. 

One can also obtain the spherical polar mode-sum expression for this Green's function:
\begin{eqnarray}
\fl
G(x,x') = \frac{1}{4\pi^{2}\alpha} \int_{0}^{\infty}\cos\omega(\tau-\tau') \rmd\omega \sum_{m=-\infty}^{\infty} \rme^{im(\phi-\phi')} \sum_{l=|m|}^{\infty} (2 \lambda+1) \frac{\Gamma(\lambda+|m|/\alpha+1)}{\Gamma(\lambda-|m|/\alpha+1)}  \nonumber\\
 P^{-|m|/\alpha}_{\lambda}(\cos\theta) P^{-|m|/\alpha}_{\lambda}(\cos\theta') R_{\omega\lambda}(r,r'),
\end{eqnarray}
where $R_{\omega\lambda}(r,r')$ is the 1D Green's function satisfying
\begin{equation}
\Big[\frac{{\rmd}}{{\rmd} r}\Big(r^{2} \frac{{\rmd}}{{\rmd} r}\Big)-\omega^{2} r^{2} -\lambda(\lambda+1)\Big]R_{\omega\lambda} = -\delta(r-r').
\end{equation}
The solution of this equation can be given in terms of Modified Bessel Functions, 
\begin{equation}
R_{\omega\lambda}(r,r') = \frac{I_{\lambda+1/2}(\omega r_{<}) K_{\lambda +1/2}(\omega r_{>})}{(r r')^{1/2}},
\end{equation}
where $I$ and $K$ are the Modified Bessel Functions of the first and second kinds, respectively.
The integral over $\omega$ can now be performed using the identity \cite{gradriz}
\begin{equation}
\int_{0}^{\infty} \cos\omega(\tau-\tau') I_{\lambda+1/2}(\omega r_{<}) K_{\lambda +1/2}(\omega r_{>})\rmd\omega = \frac{1}{2(r r')^{1/2}} Q_{\lambda}(\zeta)
\end{equation}
valid for $\lambda > -1$, where $Q_\lambda$ is the Legendre function of the second kind and
\begin{equation}
\zeta = \frac{(\tau-\tau')^{2} + r^{2} +r'^{2}}{2 r r'}.
\end{equation}
Thus, the spherical polar mode-sum expression for the Green's Function may be written as
\begin{eqnarray}
\fl
G(x,x') = \frac{1}{8 \pi^{2}\alpha r r'} \sum_{m=-\infty}^{\infty} e^{im(\phi-\phi')} \sum_{l=|m|}^{\infty} (2 \lambda+1) \frac{\Gamma(\lambda+|m|/\alpha+1)}{\Gamma(\lambda-|m|/\alpha+1)} \nonumber\\
 P^{-|m|/\alpha}_{\lambda}(\cos\theta) P^{-|m|/\alpha}_{\lambda}(\cos\theta') Q_{\lambda}(\zeta).
\end{eqnarray}

We now equate this Green's Function with the equivalent closed form expression (\ref{eq:gclosed}) to obtain the following generalized Heine Identity:
\begin{eqnarray}
\label{eq:generalizedheine}
\fl
\sum_{m=-\infty}^{\infty} \rme^{im(\phi-\phi')} \sum_{l=|m|}^{\infty} (2 \lambda+1) \frac{\Gamma(\lambda+|m|/\alpha+1)}{\Gamma(\lambda-|m|/\alpha+1)}P^{-|m|/\alpha}_{\lambda}(\cos\theta) P^{-|m|/\alpha}_{\lambda}(\cos\theta')Q_{\lambda}(\zeta)  \nonumber\\
=\frac{\sinh(\chi/\alpha)}{\sin\theta\sin\theta' \sinh\chi(\cosh(\chi/\alpha)-\cos(\phi-\phi'))}
\end{eqnarray}
where
\begin{equation}
\fl
\cosh\chi = \frac{\zeta-\cos\theta\cos\theta'}{\sin\theta\sin\theta'}\quad \Leftrightarrow \quad
\zeta = \cos \gamma + (\cosh\chi - \cos (\phi-\phi')) \sin\theta\sin\theta'.
\end{equation}
This identity, valid for $\zeta>|\cos \gamma|$, is completely analagous to (\ref{eq:heinenew}) and reduces to it in the $\alpha\rightarrow 1$ limit.


\section{Calculation of $\langle \hat{\varphi}^{2} \rangle_{ren}$}
Returning now to our calculation of the vacuum polarization on the horizon of the Schwarzschild black hole threaded by a cosmic string, we see that our generalized Heine Identity (\ref{eq:generalizedheine}) has precisely the right form to enable us to perform the mode-sum in the Green's function expression (\ref{eq:greensfn}). The Green's function, where one point has been taken to lie on the horizon, may now be written as
\begin{equation}
G(x,x')=\frac{1}{32\pi^{2} M^{2}\alpha}\frac{\sinh(\chi/\alpha)}{\sin\theta\sin\theta' \sinh\chi(\cosh(\chi/\alpha)-\cos(\phi-\phi'))}
\end{equation}
For purely radial separation, we take $\phi'\rightarrow\phi$, $\theta'\rightarrow\theta$, to yield
\begin{equation}
\label{eq:gclosedradial}
G(x,\theta;0,\theta)=\frac{1}{32\pi^{2} M^{2}\alpha}\frac{\sinh(\chi/\alpha)}{\sin^{2}\theta\sinh\chi(\cosh(\chi/\alpha)-1)}
\end{equation}
where we now have
\begin{equation}
\cosh\chi=\frac{\eta-\cos^{2}\theta}{\sin^{2}\theta}=1+ \frac{\epsilon}{M\sin^{2}{\theta}},
\end{equation}
on taking $r'=2M+\epsilon \Leftrightarrow \eta=\epsilon/M+1$.
From Eqs.(\ref{eq:gdiv}) and (\ref{eq:gclosedradial}), the renormalized vacuum polarization is given by
\begin{eqnarray}
\label{eq:phi2limit}
&&\fl\langle \hat{\varphi}^{2} \rangle_{ren}^{horizon} = \lim_{\epsilon \rightarrow 0} \Big[ G(x,0)-G_{sing}(x,0)\Big] \nonumber\\
&&\fl \ = \lim_{\epsilon \rightarrow 0} \Big[ \frac{1}{32\pi^{2} M^{2} \alpha \sin^{2}\theta} \frac{\sinh(\chi/\alpha)}{\sinh\chi(\cosh(\chi/\alpha)-1)}-\frac{1}{32\pi^{2} M \epsilon}-\frac{1}{192\pi^{2} M^{2}}\Big].
\end{eqnarray}
It is now convenient first expand in terms of $\chi$
 \begin{equation}
  \frac{\sinh(\chi/\alpha)}{\sinh\chi(\cosh(\chi/\alpha)-1)}=\frac{2\alpha}{\chi^{2}} +\Big(\frac{1}{6\alpha}-\frac{\alpha}{3}\Big) + O(\chi^{2}).
  \end{equation}
then 
  \begin{equation}
  \cosh\chi = 1+\frac{\epsilon}{M \sin^{2}\theta}\quad \Rightarrow \quad \frac{1}{\chi^{2}} = \frac{M\sin^{2}\theta}{2 \epsilon} + \frac{1}{12} +O(\epsilon^{2})
  \end{equation}
so we  obtain
\begin{equation}
\label{eq:gexpansion}
 \frac{\sinh(\chi/\alpha)}{\sinh\chi(\cosh(\chi/\alpha)-1)}=\frac{\alpha M\sin^{2}\theta}{\epsilon} +\frac{1}{6\alpha}(1-\alpha^{2}) + O(\epsilon^{2}).
 \end{equation}
Substituting (\ref{eq:gexpansion}) into (\ref{eq:phi2limit}) and taking the limit, we arrive at the renormalized vacuum polarization on the horizon,
\begin{equation}
\label{eq:phi2ren}
\langle \hat{\varphi}^{2} \rangle_{ren}^{horizon} = \frac{1}{192\pi^{2} M^{2}}\Big(1+\frac{1-\alpha^{2}}{\alpha^{2}\sin^{2}\theta}\Big).
  \end{equation}
  
  The first term here is simply the Candelas result \cite{Candelas:1980zt}, which, of course we recover when $\alpha \rightarrow 1$, and the second term represents the contribution due to the presence of the cosmic string. The result (\ref{eq:phi2ren}) has in fact been calculated by Davies and Sahni \cite{DaviesSahni} but for a restricted set of $\alpha$ values such that $(1/\alpha)$ is an integer. For this restricted class of values, the cosmic string solution is obtained as a sum of $(1/\alpha)$ images of the solution in the absence of a string. The method of images cannot be applied for general azimuthal deficits, however, so our method presents a generalization and independent derivation of the result of Davies and Sahni.
  
\begin{figure}
\centering
\includegraphics[width=11cm]{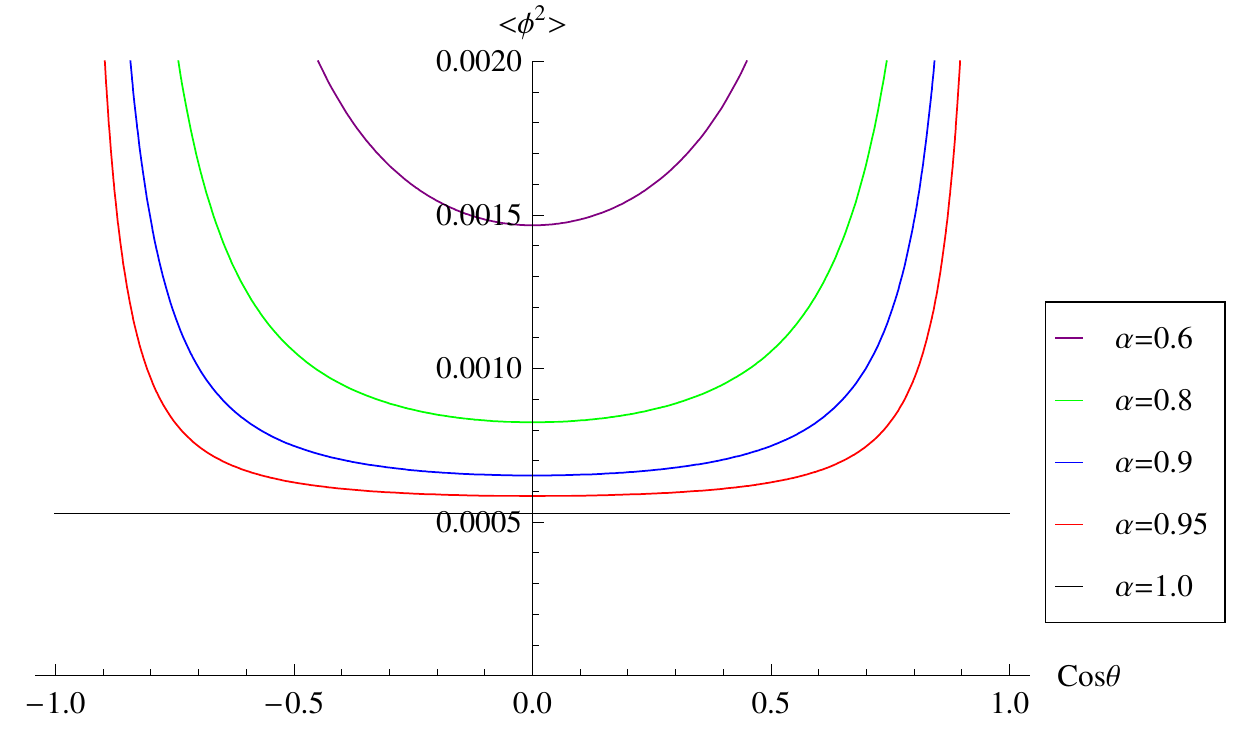}
\caption{\emph{ $\langle \hat{\varphi}^{2} \rangle_{ren}^{horizon}$} (in units $M=1$) as a function of $\cos\theta$ for a range of values of $\alpha$. For $\alpha \neq 1$,  $\langle \hat{\varphi}^{2} \rangle_{ren}^{horizon}$  diverges at the poles since there is a curvature singularity there due to the presence of the string. $\langle \hat{\varphi}^{2} \rangle_{ren}^{horizon}$ increases everywhere as $\alpha$ decreases.}
\label{fig:phi2horizon}
\end{figure}

  We have plotted the vacuum polarization in Figure~\ref{fig:phi2horizon} for a range of values of $\alpha$. We see that the presence of the string increases $\langle \hat{\varphi}^{2} \rangle_{ren}^{horizon}$ everywhere. In the equatorial plane it is increased by a factor of $1/\alpha^2$ from its value in the absence of a string.
As we approach the axis  $\langle \hat{\varphi}^{2} \rangle_{ren}^{horizon}$ diverges in a non-integrable manner as 
 \begin{equation}
\langle \varphi^{2} \rangle_{ren}^{horizon} \sim \frac{1}{48\pi^{2} }\frac{(1-\alpha^{2})}{\alpha^{2}} \frac{1}{(2M \sin\theta)^{2}} \qquad \sin\theta \to 0,
\end{equation}
  as one would expect a distance  $(2M \sin\theta)$ from a flat-space cosmic string.
  The region over which the string dominates may be characterised by the range of $\cos \theta$ (near $\theta=0$) at which 
  $\langle \hat{\varphi}^{2} \rangle_{ren}^{horizon}$ exceeds twice its value in the equatorial plane, this is given
by
  \begin{equation}
1-(\cos \theta)_2 = 1-\frac{1}{(2-\alpha^2)^{1/2}} = (1-\alpha) + O\bigl((1-\alpha)^2)\bigr) .
\end{equation}

\section{Conclusions}
We have considered  a massless, minimally coupled quantum scalar field on a Schwarzschild black hole threaded by an infinite straight cosmic string. We have obtained a simple analytic expression for the vacuum polarization on the black hole horizon for this field in the Hartle-Hawking vacuum state, which is a generalization of the Candelas \cite{Candelas:1980zt} result. In order to do this, we have derived a very useful summation formula involving non-integer Legendre functions, which we have called a generalized Heine Identity given its analogy with the standard Heine Identity involving Legendre functions of integer degree. The derivation of this formula involved equating equivalent expressions for the same Green's function on a conveniently chosen space-time. Furthermore, the derivation did not rely on specific properties of the Legendre functions. This idea can be used to derive many useful summation formulae and addition theorems for non-integer Legendre functions, some of which are presented in the Appendix.  Elsewhere \cite{CSHorizon}, we have used this idea to obtain an expression for the vacuum polarization on the entire exterior region of the space-time. We believe this framework will prove useful in the other axially symmetric black hole calculations, most importantly, the Kerr-Newman black hole. 

\section*{Appendix}
In this Appendix, we shall derive some interesting summation formulae for non-integer Legendre functions based on considering equivalent forms of the 3D Green's function on flat space threaded by a cosmic string
\begin{equation}
\rmd s^{2}=\rmd r^{2}+r^{2} \rmd\theta^{2} + \alpha^{2} r^{2}\sin^{2}\theta \rmd\phi^{2}.
\end{equation}
The 3D Green's function for Laplace's equation satisfying
\begin{equation}
\nabla^2 G_3(\mathbf{x},\mathbf{x'}) = -{g_3}(\mathbf{x})^{-1/2} \delta(\mathbf{x}-\mathbf{x'})
\end{equation}
can be written in spherical polar coordinates as the mode sum:
\begin{eqnarray}
\label{eq:app1}
\fl
G_3(\mathbf{x},\mathbf{x'})=\frac{1}{4\pi\alpha}&\sum_{m=-\infty}^{\infty} \rme^{im (\phi-\phi')}\sum_{l=|m|}^{\infty}\frac{\Gamma(\lambda+|m|/\alpha+1)}{\Gamma(\lambda-|m|/\alpha+1)}\nonumber\\ &\qquad P^{-|m|/\alpha}_{\lambda}(\cos\theta)P^{-|m|/\alpha}_{\lambda}(\cos\theta')(r_{<})^{\lambda} (r_{>})^{-(\lambda+1)} .
\end{eqnarray}
Note that this corresponds to the $n=0$ term of the 4D Green function in the corresponding ultrastatic ($g_{tt}=-1$) space-time, but that the
corresponding statement is not true in the merely static Schwarzschild cosmic string space-time.
We can also write this same Green's function in several other forms and in several coordinate systems, for example,
performing the mode decomposition in cylindrical polar coordinates, $\rho=r \sin\theta$ and $z=r\cos\theta$, we have
\begin{eqnarray}
\label{eq:app2}
\fl
G_3(\mathbf{x},\mathbf{x'})=\frac{1}{2 \pi^{2} \alpha} \sum_{m=-\infty}^{\infty} \rme^{im(\phi-\phi')}\int_{0}^{\infty} \cos k(z-z') I_{|m|/\alpha}(k\rho_{<}) K_{|m|/\alpha}(k\rho_{>}) \rmd k  .
\end{eqnarray}
The integral over $k$ may further be performed to yield
\begin{eqnarray}
\label{eq:app3}
\fl
G_3(\mathbf{x},\mathbf{x'})=\frac{1}{4 \pi^{2}\alpha (\rho \rho')^{1/2}} \sum_{m=-\infty}^{\infty} e^{im (\phi-\phi')} Q_{|m|/\alpha-1/2}\left(\frac{(z-z')^{2}+\rho^{2}+\rho'^{2}}{2 \rho \rho'}\right) .
\end{eqnarray}
Another form of this Green's function is obtained from General Axi-Symmetric Potential Theory~\cite{Weinstein, Linet1977}:
\begin{eqnarray}
\label{eq:app4}
\fl
G_3(\mathbf{x},\mathbf{x'})=\frac{1}{4 \pi^{2} \alpha}& \sum_{m=-\infty}^{\infty}\rme^{im(\phi-\phi')}\nonumber\\
 & \int_{0}^{\pi} \frac{(\rho \rho')^{|m|/\alpha} \sin^{2|m|/\alpha} \Psi }{[(z-z')^{2}+\rho^{2}+\rho'^{2}-2 \rho \rho' \cos\Psi]^{(|m|/\alpha+1/2)}} \rmd\Psi.
\end{eqnarray}
These expressions are valid for all $\alpha \in \mathbb{R}_+$; equating any one of these expressions with another gives us a formula for the functions involved. Some of these results are well known and can be found in volumes such as  Gradsteyn \& Rhyzik~\cite{gradriz}, nevertheless, we have non-obvious, important results, for example,  equating (\ref{eq:app1}) with (\ref{eq:app3}) and taking $r\rightarrow r'$, we obtain
\begin{eqnarray}
\label{eq:app5}
\fl \sum_{l=|m|}^{\infty}
\frac{\Gamma(\lambda+|m|/\alpha+1)}{\Gamma(\lambda-|m|/\alpha+1)}&& P^{-|m|/\alpha}_{\lambda}(\cos\theta)P^{-|m|/\alpha}_{\lambda}(\cos\theta')=\nonumber\\
&&
\frac{1}{\pi (\sin\theta\sin\theta')^{1/2}} Q_{|m|/\alpha-1/2}\left(\frac{1-\cos\theta\cos\theta'}{\sin\theta\sin\theta'}\right).
\end{eqnarray}

There are also a set of formulae that hold for a more restrictive set of $\alpha$ values. Linet \cite{LinetCosmicString1987} has shown that, when $\alpha > 1/2$, the Green's function may be written as
\begin{equation}
G_3(\mathbf{x},\mathbf{x'}) = \frac{1}{4 \pi \sigma}+\frac{1}{8\pi^{2}\alpha} \int_{0}^{\infty} \frac{1}{R} F_{\alpha}(u,\phi-\phi') \rmd u
\end{equation}
where
\begin{eqnarray}
\sigma&=[\rho^{2}+\rho'^{2}+(z-z')^{2}-2\rho\rho' \cos\alpha(\phi-\phi')]^{1/2} \nonumber\\
R&=[\rho^{2}+\rho'^{2}+(z-z')^{2}+2\rho\rho'\cosh u]^{1/2}
\end{eqnarray}
and
\begin{eqnarray}
\fl
F_{\alpha}(u,\Psi) = \frac{\sin(\Psi-\pi/\alpha)}{\cosh(u/\alpha)-\cos(\Psi-\pi/\alpha)}-\frac{\sin(\Psi+\pi/\alpha)}{\cosh(u/\alpha)-\cos(\Psi+\pi/\alpha)}.
\end{eqnarray}
Equating this to (\ref{eq:app1}) for $\alpha>1/2$ and taking $r\rightarrow r'$, we find
\begin{eqnarray}
\fl
& \frac{1}{\alpha}\sum_{m=-\infty}^{\infty}e^{im (\phi-\phi')}\sum_{l=|m|}^{\infty}\frac{\Gamma(\lambda+|m|/\alpha+1)}{\Gamma(\lambda-|m|/\alpha+1)} P^{-|m|/\alpha}_{\lambda}(\cos\theta)P^{-|m|/\alpha}_{\lambda}(\cos\theta')\nonumber\\
\fl &\qquad =\frac{1}{ [2(1-\cos\alpha\gamma)]^{1/2}}+\frac{1}{2\pi\alpha}\frac{1}{(2 \sin\theta \sin\theta')^{1/2}} \int_{0}^{\infty}\frac{F_{\alpha}(u,\phi-\phi')}{[\cosh\xi+\cosh u]^{1/2}} \rmd u
\end{eqnarray}
where
\begin{eqnarray}
\cos\alpha\gamma&= \cos\theta\cos\theta'+\sin\theta\sin\theta'\cos\alpha(\phi-\phi') \nonumber\\
\cosh\xi &= \frac{1-\cos\theta\cos\theta'}{\sin\theta\sin\theta'}.
\end{eqnarray}
This formula has the benefit of relating a 3D mode-sum expression to an expression that isolates the 3D Hadamard singularity structure
(the first term on the RHS) together with a regular term that depends on the boundary conditions. We have used this result to prove the regularity of the mode-sum that arises in the calculation of the vacuum polarization on the exterior region of the spacetime \cite{CSHorizon}. 

Another form of the Green's function can be given in terms of the half-integer Legendre functions by considering the Green's function in toroidal coordinates, which are related to the cartesian coordinates by
\begin{eqnarray}
\fl x=\frac{\sinh\mu \cos\phi}{\cosh\mu-\cos\eta} \qquad y=\frac{\sinh\mu\sin\phi}{\cosh\mu-\cos\eta}\qquad  z=\frac{\sin\eta}{\cosh\mu-\cos\eta},
 \end{eqnarray}
 where the ranges of the coordinates are
$ 0\le\mu <\infty$, $0\le\eta < 2 \pi$,  $0\le\phi < 2 \pi$.
The mode-sum form of the Green's function obtained by separating in these coordinates is
\begin{eqnarray}
\fl
&G_3(\mathbf{x},\mathbf{x'})= \frac{1}{4\pi^2 \alpha}\sum_{m=-\infty}^{\infty}\rme^{im(\phi-\phi')}\sum_{n=-\infty}^{\infty}e^{i n(\eta-\eta')}[(\cosh\mu-\cos\eta)(\cosh\mu'-\cos\eta')]^{1/2} \nonumber\\
\fl&\qquad
   \rme^{|m|/\alpha \pi i}\frac{\Gamma(n+|m|/\alpha+1/2)}{\Gamma(n-|m|/\alpha+1/2)}P_{n-1/2}^{-|m|/\alpha}(\cosh\mu_{<}) Q_{n-1/2}^{-|m|/\alpha}(\cosh\mu_{>}) .
\end{eqnarray}
Equating this to (\ref{eq:app3}) gives us the following addition theorem
\begin{eqnarray}
\fl
&\sum_{n=-\infty}^{\infty}\rme^{i n(\eta-\eta')}e^{|m|/\alpha \pi i}\frac{\Gamma(n+|m|/\alpha+1/2)}{\Gamma(n-|m|/\alpha+1/2)}P_{n-1/2}^{-|m|/\alpha}(\cosh\mu_{<}) Q_{n-1/2}^{-|m|/\alpha}(\cosh\mu_{>}) \nonumber\\
\fl&\qquad
=\frac{Q_{|m|/\alpha-1/2}(\chi)}{[\sinh\mu\sinh\mu'(\cosh\mu-\cos\eta)(\cosh\mu'-\cos\eta')]^{1/2}}.
\end{eqnarray}

The final coordinate system we shall investigate is prolate spheroidal coordinates which may trivially extended to the oblate spheroidal case. The relationship between prolate spheroidal coordinates and cartesian coordinates is
given by
\begin{equation}
\fl
x=\sinh\sigma\sin\theta\cos\phi \qquad y=\sinh\sigma\sin\theta\sin\phi  \qquad z=\cosh\sigma\cos\theta.
\end{equation}
where the ranges of the coordinates are $ 0\le\sigma <\infty$, $0\le\theta\le \pi$,  $0\le\phi
< 2\pi$.
The mode form of the Green's function obtained by separating in these coordinates is
\begin{eqnarray}
\fl
& G_3(\mathbf{x},\mathbf{x'}) = \frac{1}{4\pi\alpha} \sum_{m=-\infty}^{\infty} e^{i m (\phi-\phi')}\sum_{l=|m|}^{\infty}(2\lambda+1)e^{|m|/\alpha \pi i}\frac{\Gamma(\lambda+|m|/\alpha+1)^{2}}{\Gamma(\lambda-|m|/\alpha+1)^{2}}  \nonumber\\
\fl & \qquad P_{\lambda}^{-|m|/\alpha}(\cos\theta)P_{\lambda}^{-|m|/\alpha}(\cos\theta')P_{\lambda}^{-|m|/\alpha}(\cosh\sigma_{<})Q_{\lambda}^{-|m|/\alpha}(\cosh\sigma_{>})
\end{eqnarray}
We shall write down only one possible summation formula here, by equating the Green's function expression above to (\ref{eq:app3}), we obtain a summation formula for a product of four Legendre functions:
\begin{eqnarray}
\fl &
\sum_{l=|m|}^{\infty}(2\lambda+1)e^{|m|/\alpha \pi i}\frac{\Gamma(\lambda+|m|/\alpha+1)^{2}}{\Gamma(\lambda-|m|/\alpha+1)^{2}}  P_{\lambda}^{-|m|/\alpha}(\cos\theta)P_{\lambda}^{-|m|/\alpha}(\cos\theta')  \nonumber\\
\fl & \qquad P_{\lambda}^{-|m|/\alpha}(\cosh\sigma_{<})Q_{\lambda}^{-|m|/\alpha}(\cosh\sigma_{>})  
=\frac{1}{\pi\alpha} \frac{Q_{|m|/\alpha-1/2}(\chi)}{\sqrt{\sinh\sigma\sinh\sigma'\sin\theta\sin\theta'}}
\end{eqnarray}
where
\begin{equation}
\fl
\chi=\frac{\cosh^2\sigma+\cosh^2\sigma'-\sin^2\theta-\sin^2\theta'-2\cosh\sigma\cosh\sigma'\cos\theta\cos\theta'}{2\sinh\sigma\sinh\sigma'\sin\theta\sin\theta'}
\end{equation}
There are many more summation and addition formulae that can be derived in this way. Those presented here were either of particular use to us elsewhere or are to the best of our knowledge previously unpublished.

\section*{Acknowledgements}
PT is supported by the Irish Research Council for Science, Engineering and Technology, funded by the National
Development Plan. 

\section*{References}
\bibliographystyle{unsrt}

\newpage

\end{document}